\documentstyle[12pt,epsf,psfig]{article}
\begin{document}
\title{\bf  $\sigma$ Meson In $J/\Psi$ Decays}  
\author{ Wujun Huo
,
Xinmin Zhang
,  
and Tao Huang
\\
{\sl CCAST (World Lab.), P.O. Box 8730, Beijing
$100080$}\\{\sl  and}\\{\sl Institute of High Energy Physics, Academia
Sinica, P.O. Box $918(4)$},  \\{\sl Beijing $100039$, P.R.  China}
}
\date{}
\maketitle
\begin{abstract}

Recently BES at BEPC found evidence for the existence of the $\sigma$
meson in the process of
  $J/\Psi \to \sigma\omega \to
\pi\pi\omega$. 
In this paper
 we firstly discuss
  the relevant coupling $g_{\sigma\pi\pi}$
and show that the linear
 $\sigma$ model gives rise to a reasonable
description of the $\sigma$
 decay into $\pi$'s, then
 we calculate the
coupling constant 
 $g^{th}_{J/\Psi\sigma\omega}$  using the
perturbative QCD technique and the light-cone wave functions of the $\sigma$
and $\omega$ mesons.  The results show that 
 the theoretical value of
$g^{th}_{J/\Psi\sigma\omega}$ is 
 within the range of experimental value
$g_{J/\Psi\sigma\omega}$.  

\end{abstract}

 \vspace*{0.5cm} \noindent

\newpage

\section{Introduction}

Beijing Spectrometer (BES) at Beijing Electron-Positron Collider (BEPC)
recently reports an evidence for the existence of the $\sigma$ particle
in $J/\Psi$ decays. In the $\pi^+ \pi^-$ invariant mass spectrum in
the process of 
 $J/\Psi \to \pi^{+} \pi^{-} \omega$ they found a low mass
enhancement and
 the detailed analysis strongly favors that the spin-parity is
${\cal
 O}^{++}$ and the statistical significance for the existence of the
$\sigma$ particle is
 about 18 $\sigma$ \cite{bes}. The BES measured values
of the
$\sigma$ mass and width are:
\begin{eqnarray}
m_\sigma &=&390^{+60}_{-36} \mbox {\rm MeV}, \nonumber\\
\Gamma_\sigma &=&282^{+77}_{-50} \mbox {\rm MeV},
\end{eqnarray}
and the branching ratio is 
\begin{equation}
{\rm Br} (J/\Psi \to \sigma\omega \to \pi^{+} \pi^{-}
\omega)=(1.71\pm 3.4\pm 4.3)\times 10^{-3}.
\end{equation}

 The $\sigma$ particle has been absent for
many years in the Review of
Particle Physics\cite{PDG} by the Particle Data Group (PDG), however
 in the recent years there has been a revival of interest in studying 
 the light scalar-isoscalar meson, the
$\sigma$ particle, as a broad resonance\cite{shi}experimentally and
theoretically. 
 A direct experimental
evidence for the $\sigma$ meson is reported recently by the Fermilab E791 collaboration
\cite{E791} in the D-meson decay
process, $D^+
\to \sigma\pi^+ \to 3\pi$. 
Theoretically, $\sigma$ meson can play important roles in some problems. In
ref. \cite{snc}, $\sigma$ can be regarded as the dominant contribution to the
$u{\bar u} +d{\bar d}$ current. Moreover, using $sigma$ mass and width from
E791, some people\cite{sg} investage $\sigma$ contribution to $B\to 3\pi$ and
find it can explain the recent data from CLEO and BABAR collaboration.

In this paper, we study phenomenologically the decay process 
$J/\Psi \to \sigma\omega \to \pi^{+} \pi^{-} \omega$
 and discuss the couplings
$g_{\sigma\pi\pi}$ and $g_{J/\Psi\sigma\omega}$. By taking 
 the meson wavefunctions of $\sigma$ and $\omega$ to be similar to
that of $\pi$ and $\rho$, and using the perturbative QCD technique, we
calculate the decay constant
 $g^{th}_{J/\Psi\sigma\sigma}$. Our theoretical prediction is shown to be 
within the range of experimental value $g_{J/\Psi\sigma\omega}$.

\section{ Coupling constants $g_{\sigma\pi\pi}$ and 
$g_{J/\Psi\sigma\omega}$}

Given the data on the $\sigma$ mass and width measured by BES in
Eqs.(1) and (2) we study the coupling constants $g_{\sigma\pi\pi}$ and 
$g_{J/\Psi\sigma\omega}$.
For a two-body decay of particle $X$ into final states $X_1$ and $X_2$
the decay width is given by 
\begin{equation}
 \Gamma
(X\rightarrow X_1 X_2)=\frac{|{\bf p}|}{32 \pi^2 M^2_X}\int |{\cal M}|^2
d\Omega =\frac{|{\bf p}|}{8 \pi M^2_X}{|{\cal M}|^2}, 
\end{equation}
where $|{\bf p}|$ is  3-momentum of the final
state  in  the center-of-mass (c.m.)
 .
 For $\sigma \to \pi^+ \pi^-$, the $\sigma$
meson  decays $100\%$ into $\pi\pi$. Furthermore isospin conservation
requires for two thirds of the time the final states be the charged
pions. So we have:
\begin{equation}
\frac{2}{3}\Gamma^0_\sigma =g^2_{\sigma\pi\pi} \frac{1}{8\pi m^2_\sigma}
\sqrt{\frac{m^2_\sigma}{4} -m^2_\pi}.
\end{equation}
Using BES data on the $\sigma$'s mass, width, and the experimental
value of the $\pi$
mass, we obtain:
\begin{eqnarray}
  g_{\sigma\pi\pi} &=&2.0^{+0.30}_{-0.19} \,{\rm GeV}.
\end{eqnarray}
This number is surprisingly consistent with the
theoretical value 
of the linear sigma model\cite{d3pi}
\begin{equation}
g^{linear-\sigma } _{\sigma \pi \pi } = \frac{\sqrt{2} m_\sigma^2}{f_\pi}
= 1.80^{+0.50}_{-0.30} \;
{\rm GeV},
\end{equation}
where  $f_\pi$ is the pion decay constant. In Eq.(6) the $\sigma$ mass is
taken from the BES measurement in Eq.(1). 

To get the phenomenological value of $g_{J/\Psi
\sigma\omega}$, we use the full three-body decay width
(for example, see \cite{d3pi}):
\begin{eqnarray}
\Gamma (J/\Psi \rightarrow \sigma\omega \rightarrow \pi^+\pi^-\omega)
= \frac{1}{2} \frac{1}{2 m_{J/\Psi}}\; g_{{J/\Psi} \sigma \omega}^2
\; g_{ \sigma \pi \pi }^2  \nonumber\\ \int_{4 m_{\omega}^2}^{(m_{J/\Psi}
-m_{\omega})^2} \frac{d\chi^2}{2\pi}\; \; \frac{1}{8\pi}\lambda^{1/2}
\left(1,\frac{\chi^2}{m_{J/\Psi}^2},\frac{m_\omega^2}{m_{J/\Psi}^2}\right)
\nonumber\\ \times \;
\frac{1}{8\pi}\lambda^{1/2}
\left(1,\frac{m_\pi^2}{\chi^2},\frac{m_\pi^2}{\chi^2}\right)
\frac{1}{(\chi^2-m_\sigma^2)^2 + \Gamma_\sigma(\chi)^2 m_\sigma^2}
,\end{eqnarray}
where factor $1/(8\pi)\times\lambda^{1/2}$ is the phase space
integral of the corresponding two-body decay subprocess and
\begin{equation}
\lambda (a,b,c) =a^2 +b^2 +c^2 -2ab-2bc-2ca
.
\end{equation}
And in Eq. (7)
\begin{equation}
\Gamma_\sigma(\chi)\equiv \Gamma_\sigma^0 \times\left(m_\sigma /
\chi\right)\left(p^\ast(\chi)/p^\ast(m_\sigma)\right)
\end{equation}
is the
co-moving resonance width where $p^\ast(\chi) = \sqrt{\chi^2/4-
m_\pi^2}$ and $\Gamma_\sigma^0 = (370^{+60}_{-36} )$  MeV the
experimental value in Eq.(1).

With the experimental values of $g_{\sigma\pi\pi}$ in Eq.(5) and
the branching ratio in Eq.(2), solving Eq. (7) gives rise to
\begin{eqnarray}
  g^{exp}_{J/\Psi\sigma\omega} &=&7.3^{+2.6}_{-1.9}\,{\rm MeV}.  
\end{eqnarray}

\section{$g^{th}_{J/\Psi \sigma\omega}$ calculated by the perturbative QCD
 }

In this section we follow closely the calculation of the 
exclusive
decays of $\Upsilon$ in ref. \cite{chao} to study 
the decay of $J/\Psi \to \sigma\omega $. The decay amplitude
consists of two parts.
 One is the hard decay amplitude of the three-gluon modes and another is
the bound-state matrix elements of outgoing
mesons.

In general
the decay amplitude of $J/\Psi \to \sigma\omega$
can be written as
\begin{equation}
{\cal M} =\Psi_{J/\Psi} (0) \int_{-1}^{1} dx \int_{-1}^{1} dy \phi^*
_\sigma (x, M^2_c ) \phi^* _\omega (y, M^2_c ) T_h (x, y, M^2_c ),
\end{equation}
where $x =x_1 -x_2 , y =y_1 -y_2$ and $x_i$,  $ y_i$ 
are the constituent's fractional longitudinal momenta
which satisfy $\Sigma x_i =1$ and $ \Sigma
y_i =1$. In Eq.(11)
$\Psi_{J/\Psi} (0)$ is a non-relativistic approximate wave function of the
$J/\Psi$ meson, $\phi_\sigma (x, M^2_c )$ and $\phi_\omega (y, M^2_c )$
are the
distribution amplitude of the $\sigma$ and $\omega$ wave functions
respectively and $M_c$ is the charm quark mass. 

$T_h (x, y, M^2_c )$ is the hard decay amplitude of the charm quark pairs
into two light
quark-antiquark pairs, which is defined by
\begin{equation}
T_h (x , y , M^2_c )=\int\frac{d^4 l}{(2\pi)^4} {\cal T}_h (x , y , l, M^2
_c
).
\end{equation}
 There are twelve Feynmann diagrams shown in Fig. 1 contributing to
$T_h$, which are all explicitly shown in Ref. \cite{chao}. Using
the Landau rules, it
is found that  both infrared and collinear divergences exist in every diagram.
Fortunately, through a use of color neutrality and collinear Ward identities,
one can prove that when summing up all of the diagrams, the divergences
cancel
\cite{chao}.

For quarkonia $J/\Psi$,   the non-relativistic approximate
wave function $\Psi_{J/\Psi} (0)$ can be determined  by the decay  $J/\Psi
\to
 e^+ e^-$ ,
\begin{equation}
\Gamma(J/\Psi \to e^+ e^- ) = \frac{16\pi \alpha^2_s e^2}{M^2_{J/\psi}} |\Psi
(0)|^2.
 \end{equation}

According to the Brodsky-Huang-Lepage \cite{bhl}, the light-cone wavefunction of
a hadron is essentially determined by the off-shell energy variable. Thus,
for the light scalar meson, $\sigma$, we assume the light-cone wave function to
be the same  as that of  the pion \cite{huang1}
\begin{equation}
\psi_{\sigma} (x_i, {\bf k_\perp})  =
A_\sigma  \exp [-b^2_\sigma  (\frac{{\bf k^2_\perp} +m^2_u }{x_1}
+\frac{{\bf k^2_\perp} +m^2_u }{x_2})],
\end{equation}
where ${\bf k_\perp}$ is the relative transverse momentum of the final meson,
$m_u$ is $u$ quark mass and $x_i$ are the constituent's fractional
longitudinal momenta. $A_\sigma$ and $b_\sigma$ are two free parameters which
are
 taken to be $A_\sigma =A_\pi \approx 32 {\rm GeV}^{-1}$ and $b^2_\sigma
=b^2_\pi \approx 0.84
 {\rm
GeV^{-2}}$ \cite{huang1}.   For the wave function of the vector meson,
$\omega$,
we assume it to be
the same  as that of the $\rho$ meson\cite{huang2} \begin{equation}
\psi_{\omega} (y_i, {\bf k_\perp})  =
A_\omega  \exp [-b^2_\omega (\frac{{\bf k^2_\perp} +m^2_u }{y_1 }
+\frac{{\bf k^2_\perp} +m^2_u} {y_2})],
\end{equation}
where $A_\omega$ and $b_\omega$ are taken to be $A_\omega
=A_\rho \approx 30 {\rm
GeV}^{-1}$ and
$b^2_\omega =b^2_\rho \approx 0.55 {\rm GeV^{-2}}$,  which can be determined
from two constraints\cite{huang1,huang2}.

The distribution amplitude $\phi$  in Eq.(11) is defined as
\begin{equation}
\phi (x_1 , x_2, M^2_c ) =\int_{0}^{M^2_c  } \frac{dk^2_\perp}{16\pi^2} \psi
(x_1 ,x_2, k^2_\perp) .
\end{equation}
Making use of the wave function of $\sigma$ and $\omega$ in Eqs.(14) and
(15) we have: 
\begin{eqnarray}
\phi_\sigma (x, M^2_c ) =\frac{A_\sigma  (1-x^2) }{64\pi^2
b^2_\sigma } [\exp (-4b^2_\sigma \frac{m^2_u }{1-x^2})-\exp (-4b^2_\sigma
\frac{M^2_c +m^2_u }{1-x^2})],
\end{eqnarray}
\begin{eqnarray}
\phi_\omega (y, M^2_c ) =\frac{A_\omega (1-y^2)  }{64\pi^2 b^2_\omega }
[\exp (-4b^2_\omega \frac{m^2_u }{1-y^2})-\exp (-4b^2_\omega \frac{M^2_c
+m^2_u }{1-y^2})],
 \end{eqnarray}
where we have used $x = x_1 - x_2, y= y_1 - y_2$,
$\Sigma x_i =1$ and $ \Sigma
y_i =1$. In deriving the integration of Eq.(16), we have ignored the QCD
evolution on the distribution amplitude since the charm quark $M_c$ is not
large and the evolution effects are not significant in this energy range. Thus
the distribution amplitude is essentially determined by the non-perturbative
model\cite{huang1}.
  In Fig.2 we show the distribution
amplitudes of
$\sigma$, $\omega$.

With the values of the parameters $A_\pi,b^2_\pi,A_\rho,b^2_\rho$
given above
,
we obtain the coupling constant $g^{th}_{J/\Psi\sigma\omega}$ responsible for
the
decay
$J/\Psi\to\sigma\omega$ ,
\begin{eqnarray}
 g^{th}_{J/\Psi\sigma\omega}=10.7{\rm MeV}.
\end{eqnarray}
One can see that our theoretical prediction on
$g^{th}_{J/\Psi\sigma\omega}$ above
agrees within the errors with the experimental value
$g^{exp}_{J/\Psi\sigma\omega} $ in Eq.(10).

\section{Conclusion}
In this paper we have used the new BES experimental evidence for
the $\sigma$ particle and studied its properties in the process of
$J/\Psi \to \sigma\omega\to
\pi\pi\omega$. We have obtained the phenomelogical values of the
coupling constants $g_{\sigma\pi\pi}$ and $
g_{J/\Psi\sigma\omega}$, then the theoretical
prediction on the $g^{linear-\sigma}_{\sigma\pi\pi}$ in the linear
$\sigma$ model.
We show that linear $\sigma$ model gives rise to a reasonable
description of the $\sigma$ decay into $\pi$'s. We have calculated 
$g_{J/\Psi\sigma\omega}$ by the perturbative QCD and in this approach
we take
the wave
functions of the $\sigma$ and $\omega$ to be similar to that of
$\pi$ and $\rho$. Our theoretical predictions and the experimental values
of the $g_{J/\Psi\sigma\omega}$ are shown to be consistent.

\section*{Acknowledgments}

One of the authors (W. J. Huo) acknowledges  support from
the Chinese Postdoctoral  Science Foundation and CAS K.C. Wong Postdoctoral
Research Award  Fund. We thank Zhi-Peng Zheng and N. Wu for  useful
discussions.
This work is supported in part by the NSF of China.

\newpage

\begin{figure}
\vskip 5cm
\epsfxsize=20cm
\epsfysize=10cm
\centerline{
\epsffile{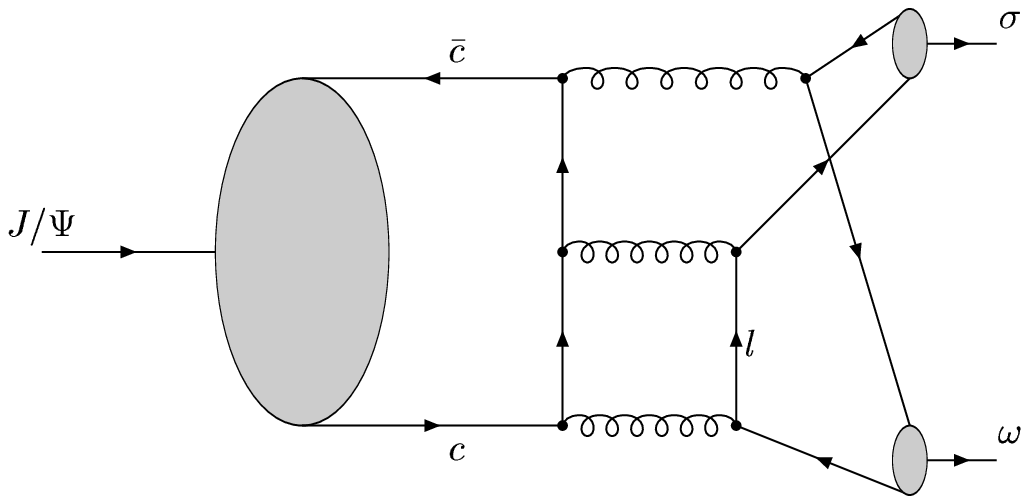}}
\vskip -8cm
\caption{Feynmann diagram for
the hard
three-gluon decay. There are 12 diagrams in total, however shown is just
one of
them. }
\end{figure}
\vskip -3cm
\begin{figure}
\epsfxsize=10cm
\epsfysize=10cm
\centerline{
\epsffile{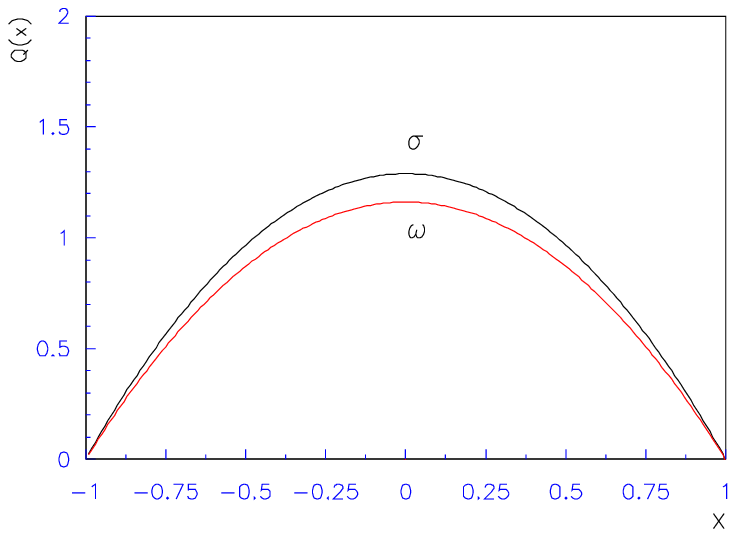}}
\vskip 0cm
\caption{Distribution amplitudes $\phi_\sigma (x)$ and $\phi_\omega (x)$ of the
$\sigma$ and $\omega$ mesons. The solid curve $\phi_\sigma (x)$ is determined by Eq. (17)
with $A_\sigma =32 {\rm GeV}^{-1}$ and $b^2_\pi =0.84 {\rm GeV}^{-2} $ and the dashed curve
$\phi_\omega (x)$ is determined by Eq. (18) with $A_\omega =30{\rm GeV}^{-1}$ and $b^2_\rho =0.55
 {\rm GeV}^{-2} $.}
\end{figure}
\end{document}